\documentstyle[aps]{revtex}

\begin{document}
\title{Mesoscopic Fluctuations Of Orbital Magnetic Response In Level-Quantized
Metals}
\author{R. A. Serota}
\address{Department of Physics\\
University of Cincinnati\\
Cincinnati, OH\ 45221-0011\\
serota@physics.uc.edu}
\maketitle

\begin{abstract}
We evaluate the distribution function of mesoscopic fluctuations of orbital
magnetic response in finite-size level-quantized metal particles and
Aharonov-Bohm rings for temperatures smaller than the mean level spacing. We
find a broad distribution with the reduced moments much larger than the
mean. For strong spin-orbit interaction we find very long tails due to
thermal activation of large effective moments of the electrons at the Fermi
level.
\end{abstract}

\section{Introduction}

The question of mesoscopic fluctuations of orbital magnetic response has
been extensively studied within the perturbation theory approach which,
generally, holds when the temperature and/or level broadening is larger than
the mean level spacing $\Delta $\cite{OZS}. In 2D (which we only consider
here), the variance of the magnetic moment was predicted to be\footnote{%
In an earlier work, Altshuler and Spivak\cite{AS} considered current
fluctuations in SNS\ junctions with broken time-reversal symmetry. This
problem is equivalent to the persistent current fluctuations in AB\ rings.
If the time-reversal symmetry is not broken and the Cooperon contribution is
taken into account, their result would be equivalent to Ref.\cite{OZS},
including the logarithmic term.},\footnote{%
Insofar as the exact numerical coefficient in eq. (\ref{M_var-pert}), both
Refs.\cite{OZS} and\cite{AS} used the Euler-Maclaurin method for Matsubara
summation with a resulting integral over the continuous variable $x=mT/E_{c}$%
, $dx=T/E_{c}\ll 1$. However, for small $x$ a more accurate procedure must
be used. It is explained in Ref.\cite{SS3} for the mean response (Appendix
II) and can be trivially extended to fluctuations.} 
\begin{equation}
\left\langle \delta M^{2}\right\rangle \sim \mu _{B}^{2}\left( k_{F}\ell
\right) ^{2}\left( \frac{\phi }{\phi _{0}}\right) ^{2}\ln \left( \frac{E_{c}%
}{T^{*}}\right) \sim \mu _{B}^{2}\left( k_{F}\ell \right) ^{2}\left( \frac{%
\mu _{B}H}{\Delta }\right) ^{2}\ln \left( \frac{E_{c}}{T^{*}}\right)
\label{M_var-pert}
\end{equation}
where

\begin{equation}
T^{*}=\max \left\{ T,\tau _{H}^{-1}\right\}  \label{Tstar}
\end{equation}
with the following notations: $\mu _{B}$ is Bohr magneton, $k_{F}$ is the
magnitude of the Fermi wave vector, $\ell $ is the electron mean-free-path, $%
\phi $ is the flux through the sample or the Aharonov-Bohm (AB) flux for
rings, $\phi _{0}=2\pi /e$ is the flux quantum in units where $\hbar =c=1$, $%
E_{c}\sim D/L^{2}$, where $D=v_{F}\ell /2$ is the diffusion coefficient, and 
$L$ the sample size (typically a ring or disk circumference), and 
\begin{mathletters}
\begin{equation}
\tau _{H}^{-1}\sim E_{c}\left( \frac{\phi }{\phi _{0}}\right) ^{2}\sim \frac{%
\left( \mu _{B}H\right) ^{2}}{\Delta }\left( k_{F}\ell \right)  \label{tau_H}
\end{equation}

The logarithmic dependence changes its form when 
\end{mathletters}
\begin{equation}
\phi \approx \phi _{c}\sim \phi _{0}\left( \frac{T}{E_{c}}\right) ^{1/2}
\label{phi_c}
\end{equation}
The significance of the flux scale $\phi _{c}$ is such that at $T\sim \Delta 
$ it determines the scale of transition from Gaussian Orthogonal (GOE)\ to
Gaussian Unitary (GUE)\ Ensemble. It is also the scale of linear response in
the problem. (While eq. (\ref{M_var-pert}) indicates that, aside from
logarithmic corrections, the linear response regime extends to $\phi \sim
\phi _{0}$ for the fluctuations, it is obtained in the so called
''zero-mode'' approximation\cite{OZS} for systems with simply connected
geometries. However, the higher modes may need to be summed up to correctly
evaluate the fluctuations\cite{GRB}. Consequently, unless specifically
stated otherwise, we shall limit our consideration to $\phi <\phi _{c}$
where the response is linear both for the fluctuations and the mean, for
simply connected and AB\ geometries.)

The average magnetic moment was predicted to be\cite{OZS} 
\begin{equation}
\left\langle M\right\rangle \sim 
{\mu _{B}\frac{\phi \phi _{0}}{\phi _{c}^{2}}\text{, }\phi <\phi _{c} \atop \mu _{B}\frac{\phi _{0}}{\phi }\text{, }\phi >\phi _{c}}%
\label{M_mean-pert}
\end{equation}
which, for $T\sim \Delta $, becomes 
\begin{equation}
\left\langle M\right\rangle \sim 
{\mu _{B}\left( k_{F}\ell \right) \frac{\phi }{\phi _{0}}\sim \mu _{B}\frac{\mu _{B}H}{\Delta }\left( k_{F}\ell \right) \text{, }\phi <\phi _{c} \atop \mu _{B}\frac{\phi _{0}}{\phi }\sim \mu _{B}\frac{\Delta }{\mu _{B}H}\text{, }\phi >\phi _{c}}%
\label{M_mean-pert-2}
\end{equation}
Comparing (\ref{M_mean-pert-2}) and (\ref{M_var-pert}) on the scale $\phi
<\phi _{c}$, we see that $\left\langle M\right\rangle $ and $\left\langle
\delta M^{2}\right\rangle ^{1/2}$ are of the same order of magnitude. While
numerically the latter is larger\cite{OZS}, it is clear that the diamagnetic
response is much less likely than the paramagnetic response.

In this work we address the regime in which level-quantization becomes
important, namely, $T\ll \Delta $. In this case, the perturbation theory
approach, which uses the level density correlation function and standard
thermodynamics, is no longer applicable and one needs to use a single
electron picture in conjunction with thermal occupancy of a two-level
system. In previous papers, we considered the mean response of GOE and
Gaussian Symplectic Ensemble (GSE), the latter being the case for strong
spin-orbit (SO)\ interaction. We have argued the for GOE\ the mean response
can be formulated in terms of a single-electron van Vleck response at the
Fermi level\cite{S1} and for GSE in terms of the effective magnetic
moments/persistent currents of electrons in the last occupied (Fermi) state%
\cite{S2}. Here, we apply these approaches to mesoscopic fluctuations.

\section{GOE}

As explained in Ref.\cite{S1}, the mean orbital response can be understood
in terms of the van Vleck (vV) response that involves virtual transitions
from the last occupied level $\varepsilon _{i}$ to the first unoccupied
level $\varepsilon _{f}$ of the Fermi sea 
\begin{equation}
\epsilon _{vV}=\frac{\left| \left\langle i\right| \widehat{M}_{z}\left|
f\right\rangle \right| ^{2}H^{2}}{\varepsilon _{i}-\varepsilon _{f}}\equiv 
\frac{\left| \widehat{M}_{if}\right| ^{2}H^{2}}{\varepsilon _{i}-\varepsilon
_{f}}  \label{E_vV-1}
\end{equation}
where $\widehat{M}_{z}$ is the magnetic moment along the magnetic field $H$
(perpendicular to the sample). There are three possible sources for
non-self-averaging (fluctuations) based on this picture. The first two,
based on eq. (\ref{E_vV-1}), are the fluctuations of $\left| \widehat{M}%
_{if}\right| ^{2}$ and the distribution of the level spacing at the Fermi
level (that is, the distribution of the values in the denominator of (\ref
{E_vV-1})). We expect the latter to be dominant and will neglect the former.
The third, which may lead to occasional diamagnetism, is the
system-dependant nature of the Fermi-sea cancellation between the
diamagnetic and paramagnetic contributions to the total magnetic response%
\cite{S1},\cite{F}; it is not studied here.

The mean value of the vV response is obtained as follows. $\left| \widehat{M}%
_{if}\right| ^{2}$ can be found using the semiclassical approximation (and,
more precisely, using the result for the magnetic dipole absorption)\cite{S1}
and is given by 
\begin{equation}
\left| \widehat{M}_{if}\right| ^{2}\sim \mu _{B}^{2}\left( k_{F}\ell \right)
\label{M_if-est}
\end{equation}
In what follows, we will use a dimensionless measure $x$ for the level
spacing at the Fermi level, 
\begin{equation}
\varepsilon _{f}-\varepsilon _{i}=x\Delta  \label{x}
\end{equation}
Averaging with the GOE \cite{BFFMPW} distribution function for the nearest
energy levels, we find\cite{S1} 
\begin{eqnarray}
\overline{\epsilon _{vV}} &=&-s\left| \widehat{M}_{if}\right| ^{2}H^{2}\frac{%
\pi }{2\Delta }\int_{0}^{\infty }\exp \left( -\frac{\pi x^{2}}{4}\right) dx=-%
\frac{\pi \upsilon }{2}\left| \widehat{M}_{if}\right| ^{2}H^{2}
\label{E_vV-GOE} \\
&\sim &-\left| \chi _{L}\right| H^{2}A\left( k_{F}\ell \right) \sim -\tau
_{H}^{-1}  \label{E_vV-tau_H}
\end{eqnarray}
Here $s$ is the level degeneracy ($s=2$, on the account of spin), $\upsilon
=s\Delta ^{-1}$ is the density of levels, $\chi _{L}$ is the Landau
susceptibility\cite{LL} and $A$ is the sample area. As was pointed out
earlier, it is assumed that $\tau _{H}^{-1}\ll \Delta $, with the opposite
limit corresponding to GUE.

To consider the fluctuations, we notice that the second order perturbation
theory is used in derivation of (\ref{E_vV-1}). Consequently, the limit of
its applicability\cite{LL2} is, per (\ref{E_vV-1}) and (\ref{x}), 
\begin{equation}
\epsilon _{vV}\sim \frac{\tau _{H}^{-1}}{x}\ll x\Delta  \label{E_vV-2}
\end{equation}
that is 
\begin{equation}
x\gg \left( \tau _{H}^{-1}/\Delta \right) ^{1/2}\text{ or }\varepsilon
_{f}-\varepsilon _{i}\gg \left( \tau _{H}^{-1}\Delta \right) ^{1/2}
\label{x-cond}
\end{equation}
An important observation about the above conditions is that the range of
linear response is sample-specific and depends on the Fermi level spacing of
a particular particle. On the other hand, when describing the statistical
distribution of the magnetic energies, it is only meaningful to consider
their values at a given field (or a range of fields).

An estimate of the fluctuation can be obtained by substituting the r.h.s. of
(\ref{x-cond}) into eq. (\ref{E_vV-1}) 
\begin{equation}
\epsilon _{vV}\sim \left( \tau _{H}^{-1}\Delta \right) ^{1/2}\gg \overline{%
\epsilon _{vV}}  \label{flct}
\end{equation}
Below we will show that (\ref{flct})) corresponds to the result for the
reduced higher cumulants for the distribution of magnetic energies and\ is
due to the systems with small level spacing at the Fermi level.

Since the reduced moments grow with the order of the moment, the cumulants
and the moments should be of the same order of magnitude and it is
sufficient to evaluate the latter. We begin with the evaluation of the
variance/second moment of the vV response 
\begin{equation}
\overline{\epsilon ^{2}}\sim \frac{\pi }{2}\tau _{H}^{-2}\int_{\left( \tau
_{H}^{-1}/\Delta \right) ^{1/2}}\frac{dx}{x}\exp \left( -\frac{\pi x^{2}}{4}%
\right) \sim \tau _{H}^{-2}\ln \left( \frac{\Delta }{\tau _{H}^{-1}}\right)
\label{var}
\end{equation}
(omitting subscript ''$vV$''). The higher moments and the reduced moments
can be evaluated similarly and are given by 
\begin{eqnarray}
\overline{\epsilon ^{n}} &\sim &\frac{\pi }{2}\tau _{H}^{-n}\int_{\left(
\tau _{H}^{-1}/\Delta \right) ^{1/2}}x^{-n+1}dx\exp \left( -\frac{\pi x^{2}}{%
4}\right)  \label{moments} \\
^{n}\sqrt{\overline{\epsilon ^{n}}} &\sim &\left( \tau _{H}^{-1}\Delta
\right) ^{1/2}\left( \frac{\tau _{H}^{-1}}{\Delta }\right) ^{1/n}\stackrel{%
n\rightarrow \infty }{\rightarrow }\left( \tau _{H}^{-1}\Delta \right) ^{1/2}
\label{reduced_moments}
\end{eqnarray}
in agreement with the estimate (\ref{flct}). Alternatively, eqs. (\ref{var}%
)-(\ref{reduced_moments})\ can be evaluated using the distribution function
for the fluctuations which is found as 
\begin{equation}
P\left( \epsilon \right) =\frac{\pi }{2}\int_{\left( \tau _{H}^{-1}/\Delta
\right) ^{1/2}}dx\delta \left( \epsilon -\frac{\tau _{H}^{-1}}{x}\right)
x\exp \left( -\frac{\pi x^{2}}{4}\right) =\frac{\pi }{2}\frac{\tau _{H}^{-2}%
}{\epsilon ^{3}}\exp \left( -\frac{\pi \tau _{H}^{-2}}{4\epsilon ^{2}}%
\right) \text{, }\epsilon <\left( \tau _{H}^{-1}\Delta \right) ^{1/2}
\label{distr_funct}
\end{equation}
with the case $\epsilon >\left( \tau _{H}^{-1}\Delta \right) ^{1/2}$ (or,
equivalently, $x<\left( \tau _{H}^{-1}/\Delta \right) ^{1/2}$) addressed
below.

A comparison should be made between (\ref{var}) and the perturbative result (%
\ref{M_var-pert}). First, assuming for the latter that $T\sim \Delta $ (the
limit of applicability of the perturbative approximation), we find that the
corresponding energy fluctuation is 
\begin{equation}
\overline{\epsilon _{pert}^{2}}\sim \tau _{H}^{-2}\ln \left( \frac{E_{c}}{%
\Delta }\right)  \label{var_pert}
\end{equation}
The difference in $\log $ factors is because (\ref{var}) is obtained in a
two-level approximation, whereas (\ref{var_pert}) takes into account
contribution of levels within $E_{c}$ of the Fermi level; otherwise, the
results are in qualitative agreement. (The mean values $\overline{\epsilon
_{vV}}$ are also in agreement\cite{S1})$.$ Although the higher moments had
not been evaluated perturbatively, it is expected that $^{n}\sqrt{\overline{%
\epsilon _{pert}^{n}}}\sim \tau _{H}^{-1}$ and are smaller than obtained in
a two level picture. This should be anticipated since the two-level response
is very sensitive to the variations of the energy level spacing at the Fermi
level.

For $x<\left( \tau _{H}^{-1}/\Delta \right) ^{1/2}$ the perturbation theory
evaluation of $\epsilon _{vV}$ (\ref{E_vV-1}) is no longer valid. This
applies to the fraction $\sim \tau _{H}^{-1}/\Delta $ of all systems with
sufficiently small Fermi level spacing. There are two possible approaches to
this case that yield, essentially, the same result. In the first approach,
the Fermi level and the first unoccupied state can be viewed as,
effectively, a doubly degenerate state with respect to the magnetic field
perturbation. Using the secular equation\cite{LL2}, we find that the
magnetic field will split the levels in the amount 
\begin{equation}
\sqrt{\left| \widehat{M}_{if}\right| ^{2}H^{2}}\sim \left( \tau
_{H}^{-1}\Delta \right) ^{1/2}  \label{degenerate}
\end{equation}
(The second order term is still given by eq. (\ref{E_vV-1}) but with the
final state $f$ being the next unoccupied state at a distance greater than $%
\left( \tau _{H}^{-1}\Delta \right) ^{1/2}$ from the Fermi level.).
Alternatively, one can argue that such systems are effectively in the GUE\
regime. The perturbative expression for the mean value is 
\begin{equation}
\overline{\epsilon _{pert\text{ }\left( GUE\right) }}\sim \left( \tau
_{H}^{-1}\Delta \right) ^{1/2}  \label{E_pert-GUE}
\end{equation}
which is obtained from the second eq. (\ref{M_mean-pert-2}) using the
substitution $\Delta \rightarrow \left( \tau _{H}^{-1}\Delta \right) ^{1/2}$
, where the latter is the upper bound of the level spacing in the systems
that have effectively undergone the $GOE\rightarrow GUE$ transition.

The physical interpretation of the above results may be as follows. In a
particular systems, the linear response regime is determined by the Fermi
level spacing in this system, as per eq. (\ref{E_vV-2}), 
\begin{equation}
M\sim \frac{\mu _{B}^{2}H\left( k_{F}\ell \right) }{x\Delta }  \label{M_ind}
\end{equation}
As a result of linear response, the magnetic moment approaches the value 
\begin{equation}
M\sim \mu _{B}\left( k_{F}\ell \right) ^{1/2}  \label{M_max}
\end{equation}
with little change for larger fields (until $\tau _{H}^{-1}$ approaches $%
\Delta $ - see below). The statistical distribution of the response implies
the need to consider the same range of fields for all systems. The mean
magnetic moment is given by 
\begin{equation}
\left\langle M\right\rangle \sim \frac{\mu _{B}^{2}H\left( k_{F}\ell \right) 
}{\Delta }\stackrel{\tau _{H}^{-1}\sim \Delta }{\rightarrow }\mu _{B}\left(
k_{F}\ell \right) ^{1/2}  \label{M_typ}
\end{equation}
The slowly decaying distribution function 
\begin{equation}
P\left( \epsilon \right) \propto \frac{\tau _{H}^{-2}}{\epsilon ^{3}}\text{, 
}\tau _{H}^{-1}<\epsilon <\left( \tau _{H}^{-1}\Delta \right) ^{1/2}
\label{distr_funct-short}
\end{equation}
is due to systems with small Fermi level spacing, where the magnetic moment
may become as large as (\ref{M_max}).

At $\tau _{H}^{-1}>\Delta $, the single-level ansatz is no longer valid and
the results of Ref.\cite{OZS} (as described in the Introduction) should be
applied instead; at $\tau _{H}^{-1}\sim \Delta $ the latter are consistent
with the present results.

\section{GSE}

It was argued\cite{KZ},\cite{S2} that in a GSE the electrons at the Fermi
level have a magnetic moment 
\begin{equation}
M\sim \mu _{B}\left( k_{F}\ell \right) ^{1/2}  \label{M_GSE}
\end{equation}
(which is, incidentally, of the same order of magnitude as given by (\ref
{M_max})) induced by the spin-orbit interactions. For a system with even
number of electrons, the magnetic moments of the two Fermi-level electrons
cancel each other. A system with an odd number of electrons, on the other
hand, should have a permanent magnetic moment\cite{S2} and a Curie-like
susceptibility. This is in incomplete analogy with the purely spin magnetism%
\cite{SS2}, the sole difference being that $M=\mu _{B}$ in the latter case.
Consequently, we refer to Ref.\cite{S2} for the discussion of the
distribution function of the susceptibility obtained in the two- and
three-level approximation for the even- and odd-electron systems
respectively; in both cases, one finds long tails due to thermal activation
of the magnetic moments.

It should be noted that the single-electron magnetic moment (\ref{M_GSE})
can be attributed to the large electron $g$-factor, $g\sim \left( k_{F}\ell
\right) ^{1/2}$, induced by the spin-orbit interaction\cite{S2},\cite{ZS},%
\cite{MGL}. Furthermore, it was argued\cite{ZS},\cite{MGL} that the latter
can fluctuate by as much as the order of magnitude. We point out, however,
that in the temperature regime $T\ll \Delta $ considered here, the thermal
activation effects should be dominant as they may produce much larger
fluctuations.

\section{Discussion}

The two-level picture considered here predicts the large fluctuations of the
orbital magnetic response due to the possibility of small level separations
at the Fermi level. Obviously, the possible observation of these effects
imposes restrictions on the temperature and the magnetic field. Moreover,
such restrictions are sample specific since the magnetic energy and the
temperature must be compared with a particular Fermi level separation of the
nearest states. (See, for instance, eqs. (\ref{E_vV-2}) and (\ref{x-cond})).
On the other hand, a statistical description of the fluctuations implies a
common (range of) magnetic field and temperature for all systems.
Consequently, while in a given system one can observe the linear response
regime due to the vV energy (\ref{E_vV-1}) by sufficiently reducing the
magnetic field, the Fermi-level spacing in this system may be smaller than
the magnetic energy corresponding to the field at which {\it all} systems
are analyzed. The latter is critical for understanding of the effective
cut-offs in the analysis of the distribution function.

The issues that remain to be understood are the fluctuations of the matrix
elements $\left| \widehat{M}_{if}\right| ^{2}$ and the details of the
cancellation between the van Vleck paramagnetism and the precession
diamagnetism over the Fermi sea. These will require further, largely
numerical, studies. Also, our formalism should be applicable, with minor
changes, to orbital magnetism of integrable systems, such as a rectangle
with incommensurate sides\cite{vRvL}. We hope to address this problem in a
future work.

\section{Acknowledgments}

I wish to thank Bernie Goodman for many helpful discussions. This research
was not supported by any funding agency.

\end{document}